# Valence and conduction bands engineering in halide perovskites for solar cell applications

Simone Meloni[†], Giulia Palermo[†], Negar Ashari Astani[†], Basile F.E. Curchod[†], Michael Graetzel[‡], Ursula Roethlisberger[†].

† Laboratory of Computational Chemistry and Biochemistry, Institut des Sciences et Ingénierie Chimiques, École Polytechnique Fédérale de Lausanne, CH-1015 Lausanne, Switzerland

‡ Laboratoire de Photonique et Interfaces, Institut des Sciences et Ingénierie Chimiques, École Polytechnique Fédérale de Lausanne, CH-1015 Lausanne, Switzerland

*Supporting Information Placeholder*

**ABSTRACT:** we performed *ab initio* simulations aimed at identifying the atomistic and electronic structure origin of the high valence and conduction band, and band gap tunability of halide perovskites. We found that the two key ingredients are the overlap between bivalent cation and halide atomic orbitals, and the electronic charge of the bottom of the conduction band (CBM) state on the Sn or Pb atoms. In particular, we found that lower gaps are associated to higher negative antibonding overlap, and higher CBM charge on the bivalent cation. Both overlap and CBM charge on Sn/Pb can be tuned by the chemical nature of halide, bi and monovalent cation, as well as the symmetry of crystal structure. On the basis of our results we provide some practical rules to tune the valence band maximum, conduction band minimum, and band gap in this class of materials.

Mixed Organic and fully inorganic halide perovskites, $ABX_3$ (A = organic or inorganic monovalent cation, B = bivalent cation, X = halogen – See Figure 1), recently emerged as very efficient light harvesting materials for solar cells.[1-3] Power Conversion Efficiencies (PCE - percentage of solar light converted into electric current) as high as 17.9% (certified) have been reported for perovskite based solar cells.[4]

Several strategies have been attempted to increase the PCE of halide perovskites, some targeting the quality of perovskite films, e.g. its uniformity,[1] others focusing on their intrinsic optical and electronic properties, such as tuning of the band gap, $E_g$.[5-7] In the latter case, the objective is obtaining a compound with a band gap close to the ideal value of 1.1 eV.[8] Seok and Coworkers have shown that in mixed I/Br perovskites, $ABI_{3-x}Br_x$, $E_g$ can be tuned by changing the halide composition, $x$.[5] In particular, in Methyl Ammonium Lead Iodine/Bromine perovskite, $MAPbI_{3-x}Br_x$, (MA = Methyl Ammonium), the band gap widens with $x$. Kanatzidis and Coworkers [9] studied the dependence of the valence band maximum (VBM) and conduction band minimum (CBM) with chemical composition, and concluded that in $MASnI_{3-x}Br_x$ the widening of the band gap with the amount of Br is due to an increase of the energy of the latter rather than a reduction of the energy of the former.

Snaith and Coworkers confirmed the findings of Ref. [5], and shown a correlation between $E_g$ with the pseudocubic lattice parameter of the crystalline sample.[7] This relation inspired them to use the size of the monovalent cation, A, to tune $E_g$. An analogous strategy has been developed in parallel by Grätzel and Coworkers,[3] who also explored the effect of using a mixture of MA and Formamidinium (FA). None of these studies have investigates the "active" role of A on the electronic structure of the system, e.g. *via* the formation of hydrogen bonds or other strong interactions with the halides atoms of the framework.

The effect of cation and structure of the PbI framework of perovskites (e.g. the X-B-X angle) on $E_g$ has also been discussed by Amat et al. [10] They concluded that larger cations, which favor a more cubic-like structure of perovskites, red-shift the band gap. This result is somewhat counterintuitive at the light of results of Ref. [7], where it is shown that FASnBr$_3$, which has a cubic structure and smaller lattice, has a larger band gap than FASnI$_3$, which is tetragonal and has a larger lattice. This apparently conflicting results suggest that there is a significant interplay between the chemistry and structure of perovskite, and the combined effect of these two parameters on the band gap is non trivial.

The intricacy of the relation between chemical composition, crystal structure and properties of perovskites call for a comprehensive theory. This should aim at rationalizing experimental and computational results in terms of the effect of variables (chemical composition, lattice size, tilting angle) on the electronic structure of perovskite, and thus on their optical properties. This is the objective of the present work. We consider a wide range of perovskites, differing in the chemical nature of the bivalent (Pb and Sn) and monovalent (Na, Li and Cs) cations, halogen composition (I, Br, Cl), and crystal symmetry (cubic, tetragonal and orthorhombic). We also considered lead-iodide perovskites of NH$_4^+$ and PH$_4^+$ to investigate the effect of the strength of hydrogen bond with atoms of the framework on the electronic structure of the system.

We perform *ab initio* calculation and analyze the structure of VBM and CBM in terms of atomic orbitals and give a simple and intuitive interpretation of our and literature results. We also show that strategies for tuning VBM and CBM energy levels, and the bandgap must be crystal symmetry-dependent: they are different for cubic and tetragonal/orthorhombic structures.

## Results

Halide perovskites are made by corner sharing BX$_6$ octahedra forming cuboctahedral cavities, which are occupied by the A ions. In the cubic crystal, BX$_6$ polyhedra are oriented such that all B-X-B angles are 180° (Figure 1A). In Tetragonal (Figure 1B) and orthorhombic (Figure 1C) crystals the BX$_6$ polyhedra are rotated (tilted) with respect to their orientation in the cubic structure. The tetragonal structure is characterized by only one tilting angle different from 0, while in the orthorhombic structure all of them are non zero.

The chemical nature of halides and bivalent cation affects mainly the lattice size (Figure 1D), as expected on the basis of well-established empirical relations between their ionic radii and the perovskite lattice size.[11] The effect of substituting the monovalent cation, on the other hand, depends on the crystalline symmetry of the reference system. If the system is cubic, the replacement of the original cation results in a corresponding change of the lattice parameter of the crystal: if the new cation is bigger the lattice expands, if it is smaller the lattice shrinks (Figure 1E). The value of tilting angles $\vartheta_1, \vartheta_2$ and $\vartheta_3$, which measure the rotation of BX$_6$ octahedra around their three main axes with respect to the orientation in the cubic case (see Figure 1), is 0°. If the original system is tetragonal or orthorhombic, the substitution of A alters both the size of the lattice and the tilting angles. The effect of substitution on the lattice size is analogous to the cubic case: larger cations expand the lattice, and *vice versa*. The tilting angle(s), instead, increases if the new cation is smaller, and decreases if it is bigger *(see Figures 1G and 1E)*. This affects the linearity of B-X-B: the more the structure is tilted the more the B-X-B angles depart from the 180° value of cubic structures. We observed values of $\vartheta$ in the range ~7-19.5° for the non-zero tilting angles in the tetragonal structures, and ~6.5-23° in the orthorhombic ones (all tilting angles are non zero, in this latter case). Special is the case of monovalent cations forming hydrogen bonds with the halides of the framework. Depending on the strength of the hydrogen bond, the cubic structure might or might not be a (meta)stable state. In fact, the structure obtained by replacing Cs with NH$_4$ in the cubic CsPbI$_3$ is triclinic, with crystallographic angles $\alpha = \beta = 85.8°$ and $\gamma = 87.1°$ (Figure 1H). If we replace, instead, Cs with a weaker hydrogen bond donor cation, e.g. PH$_4$, we obtain a stable cubic structure (Figure 1I). Moreover, strong hydrogen bonds induce deformations of BX$_6$ polyhedra (Figure 1H).

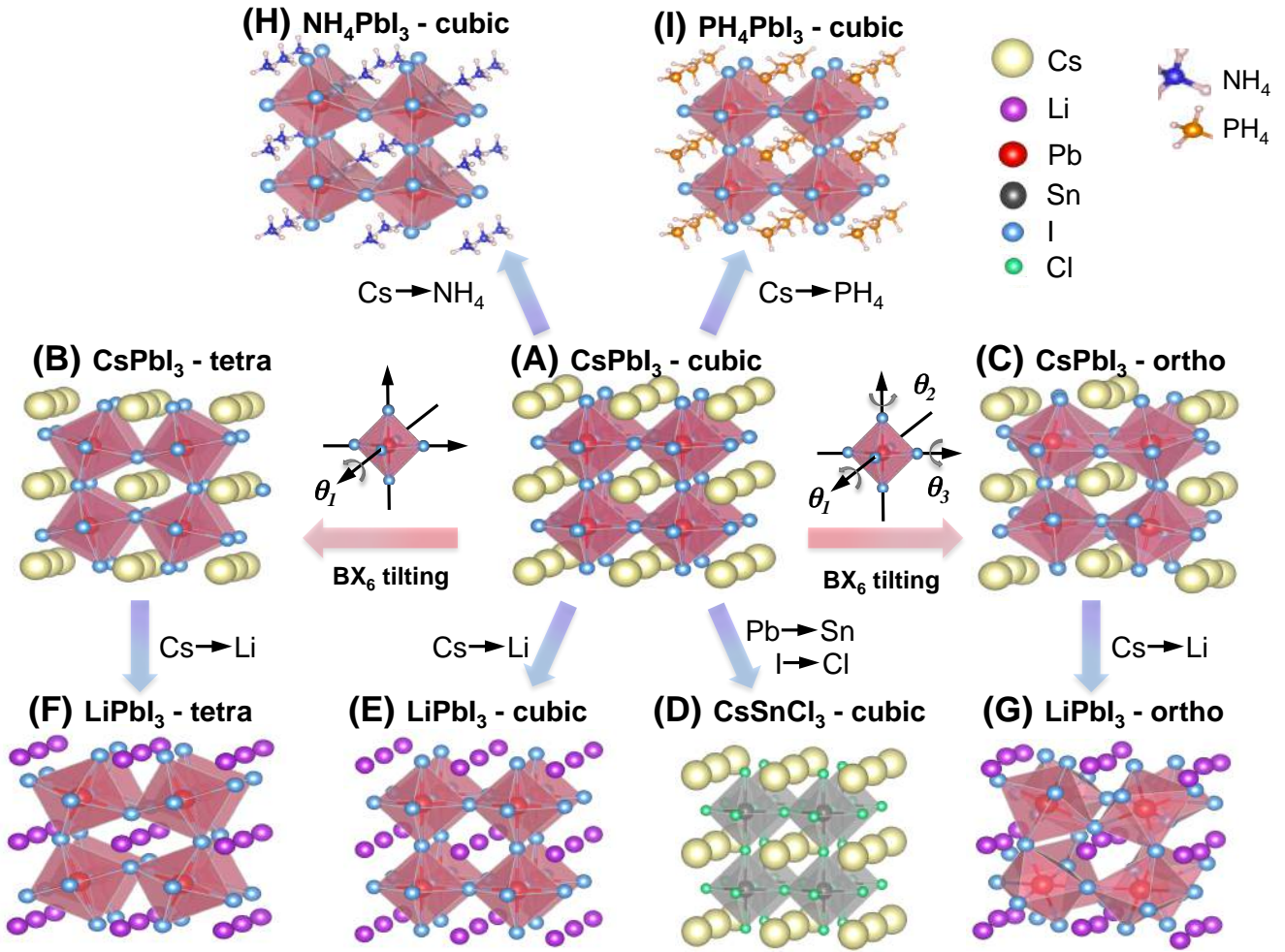

**Figure 1:** Configurations of the equilibrium structure of several perovskites. Images are ordered such that the corresponding structures can be though as a series of alteration of the CsPbI$_3$ cubic structure (A), characterized by a lattice parameter $a = 6.38$ Å. In B and C are shown the tetragonal and orthorhombic structures of CsPbI$_3$, respectively. They can be though as obtained form the cubic analogue by tilting the PbI$_6$ octahedra along their axis parallel to the tetragonal axis (tetragonal structure) or along all of their three axes (orthorhombic). Tetragonal CsSnI$_3$ is characterized by a tilting angle $\vartheta_1 = 14.3°$ and a pseudocubic lattice parameter $a^* = \sqrt[3]{V} = 6.12$ Å ($V$ is the volume of the unit cell). In the orthorhombic CsSnI$_3$ $\vartheta_1 \sim \vartheta_3 \sim \vartheta_3 = 8.5°$ and $a^* = 6.30$ Å. (D) is obtained from the cubic CsPbI$_3$ by replacing Pb with Sn and I with Cl ($\vartheta_1 = \vartheta_3 = \vartheta_3 = 0°$ and $a = 5.85$ Å). E, F and G are obtained from the cubic, tetrahedral and orthorhombic CsPbI$_3$, respectively, by replacing Cs with Li. Their tilting angle and pseudocubic lattice parameters are: E) $a = 6.32$ Å; F) $\vartheta_1 = 20.3°$, $a = 6.12$ Å; G) $\vartheta_1 = 20.5°$, $\vartheta_2 = 18.8°$, $\vartheta_3 = 10.8°$, $a = 5.81$ Å. (H) and (I) are obtained by replacing Cs in the cubic CsPbI$_3$ with NH$_4$ and PH$_4$, respectively. While PH$_4$PbI$_3$ preserves the original cubic symmetry ($a = 6.32$ Å), NH$_4$PbI$_3$ is triclinic ($\alpha = \beta = 85.8°$, $\gamma = 87.1°$) and the PbI$_6$ octahedra are highly distorted ($a = 6.38$ Å).

An interesting characteristic of all the perovskites considered in this work is the persistence of the features of their VBM and CBM orbitals. VBM is an antibonding combination of B n$s$ and X m$p$ orbitals, namely Sn-5$s$ or Pb-6$p$, and Cl-3$p$, Br-4$p$ or I-5$p$ (see Figure 2). This orbital has a high covalent character, with a typical B/X atomic orbitals contribution of ~30-40/70-60%, depending on the chemical nature of A, B and X, and the crystal symmetry. CBM is also an antibonding combination, of B n$p$ and X m$s$ orbitals this time (see also Ref. [12]). CBM is more ionic, with B orbitals contributing ~70-90% to it.

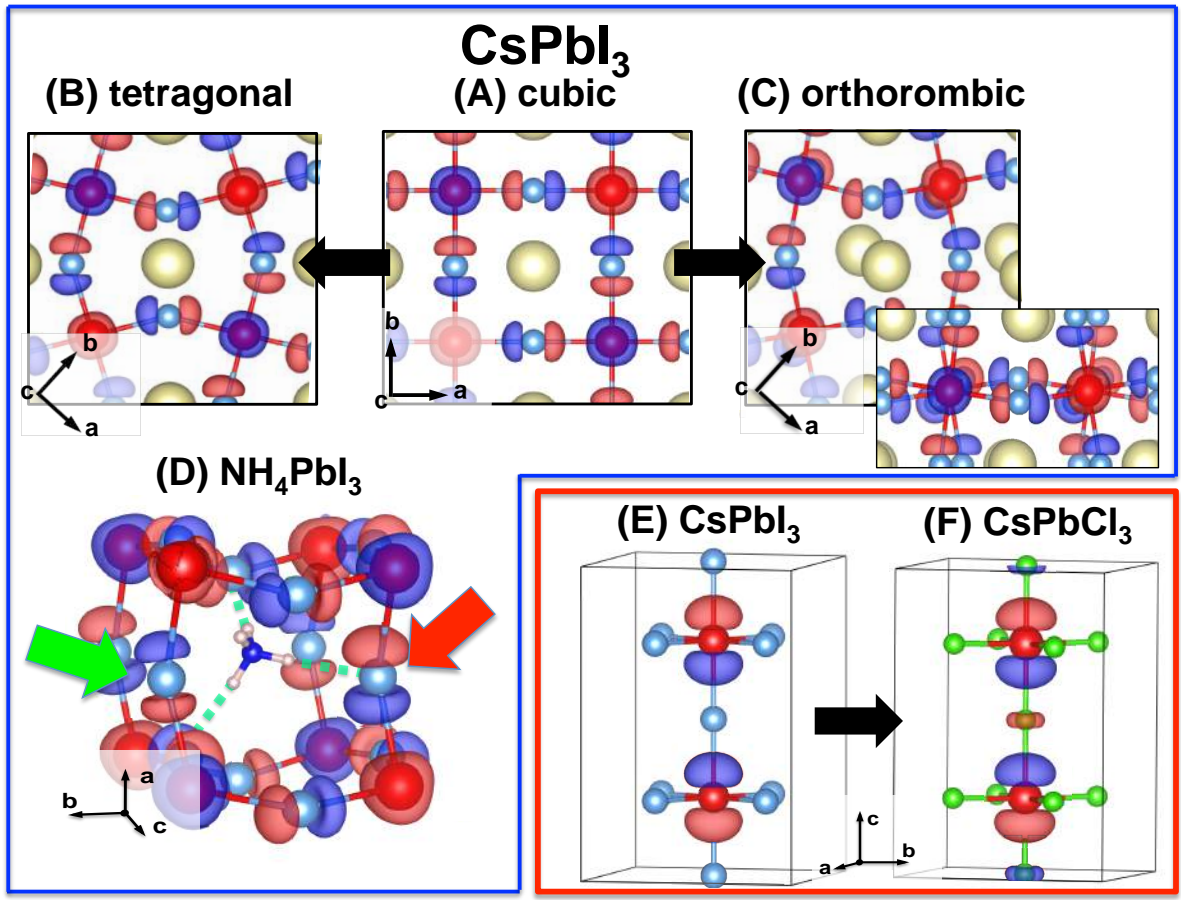

Figure 2: VBM (blue frame) and CBM (red frame) orbitals of selected systems. Panels (A), (B) and (C) shows how the tilting of $BX_6$ octahedra affects the overlap: in the cubic structure (A) X $mp$ orbitals are aligned with B-X bonds and the absolute value of the negative overlapping is maximum ($O_{VBM} = -0.21, E_{VBM} = -0.02$ eV). In tetrahedral (B) and orthorhombic (C) structures $p$ orbitals are no longer well aligned, and the overlap diminishes (tetrahedral $O_{VBM} = -0.17, E_{VBM} = -0.31$ eV; orthorhombic $O_{VBM} = -0.11, E_{VBM} = -0.56$ eV). In the case of orthorhombic structures, the tilting along all the three of the octahedra further reduce the overlap, as shown from the lateral view of the crystal shown in the inset of panel (C). (D) Monovalent cations forming hydrogen bonds (green dashed lines) distort the framework reducing the overlap with the same mechanism of tilting angles. In addition, hydrogen bonds polarize the VBM orbitals, resulting in an increase of electronic density on hydrogen bond (red arrow) and a reduction on the others (green arrow) and on bivalent cations. (E) and (F) CBM orbital in tetragonal $CsPbI_3$ and $CsPbCl_3$. The effect of halide substitution along the series I, Br, Cl is moving the CBM electronic charge from B to X. For example, $q_B = 0.81$ and $E_{CBM} = 0.76$ eV in $CsPbI_3$, and $q_B = 0.86$ and $E_{CBM} = 0.34$ eV in $CsPbI_3$. A similar effect is produced by the change of crystal structure along the series cubic to tetragonal to orthorhombic, and monovalent cation from low to high charge density ions.

To understand the effect of all the alterations considered in experiments, we have investigated the effect of variation of the chemical nature of A, B and X, and crystalline symmetry on the VBM and CBM. It is worth remarking that simple hypotheses to explain the phenomenology observed in experiments, such us that the energy of VBM and CBM changes because the energy of the corresponding atomic orbitals changes with the nature of the halide, are inadequate. Our calculations show that the difference between the VBM and CBM energy of perovskites of different halides change in a range of ~2 eV (see Figure SI1), and sometimes have an opposite sign with respect the difference of energy between I, Br and Cl $p$ atomic orbitals ($\Delta E_{I/Br} \sim \Delta E_{Br/Cl} = 0.5$-$0.6$ eV). Thus, it is clear that other effects, discussed in the following, are responsible for the tunability of the valence and conduction bands, and, thus, band gap of halide perovskites.

Concerning the VBM state, the effect of changing symmetry, from cubic to tetragonal to orthorhombic, is reducing the overlap between B $ns$ and X $mp$ orbitals (see Figure 2/A, B and C). In the cubic structure X $mp$ orbitals are aligned along the B-X bond. This alignment is worse in the other two structures; in particular it decreases along the series cubic → tetragonal → orthorhombic. As for monovalent cation substitution, we have an effect that depends on the symmetry of the crystal. In cubic systems, in which the size of A affects only the size of the lattice, we observe an increase or decrease of the (negative) B-$ns$/X-$mp$ overlap with the size of the cation. In tetragonal and orthorhombic crystals, in which the size of A affects both the lattice size and the tilting angle, we have two competing effects: increase or decrease of the B-n$s$/X-m$p$ overlap due to i) size of the lattice and ii) tilting angle. The overlap increases with the shrinking of the lattice, like in the case of cubic systems, and decreases with the increase of $\vartheta$. Simulations show that, typically, the second effect dominates, and the overlap is reduced in systems with smaller A. The chemical nature of A can affect the VBM orbital also in a more direct way, by polarizing it. This occurs when A can establish a relative strong bond with the halide of the framework, like hydrogen bonds (Figure 2/D). The effect is twofold in this case. First, because the structure is highly distorted, the alignment of X-m$p$ orbitals with B-X bonds is very poor. Second, the electron density on the X atom hydrogen bonded with A is higher, with corresponding charge depletion on the B and the remaining X atoms. Both effects cooperate at reducing the negative B-n$s$/X-m$p$ overlap. Finally, concerning the effect of substituting the halide, we mentioned that the change of the chemical nature of this species affects the geometry of the system by shrinking/expanding the lattice. However, the size of the $p$ orbitals of X do not shrinks/expands of the same amount, thus producing a change of the antibinding overlap with the halide composition of the perovskite.

Summarizing, the main effect of all the changes discussed is altering the overlap between the B n$s$ and X m$p$ atomic orbitals forming the VBM. According to the tight binding formulation, the energy of a crystal orbital depends on the overlap among the relevant atomic orbitals. In particular, the larger is the negative (antibonding) overlap the higher is the energy of the state. This argument, together with the above analysis, suggests that the key observable correlating with $E_{TVB}$ is the overlap, and A, B and X substitutions, or the crystal symmetry are all effective but equivalent ways to alter the B n$s$/X m$p$ overlap. To validate this idea we computed the overall overlap in the VBM orbital, $O_{VBM} = Re(\sum_{i \in X-mp} \sum_{j \in B-ms} c^*_{VBM,i} c_{VBM,j} \mathfrak{O}_{ij})$, with $\mathfrak{O}_{ij} = \int dr\, \phi^*_i(r)\phi_j(r)$ overlap between pairs of X and B atomic orbitals, and $c_j$ and $c_i$ projection coefficients of crystal orbitals over the atomic ones. In Figure 3 we show $O_{VBM}$ vs $E_{VBM}$. The fitting of this correlation with a linear function is pretty good for Sn-based perovskites (regression coefficient $R^2 \sim 0.93$), while it is poorer for Pb-based systems ($R^2 \sim 0.52$). However, if we focus on Pb perovskites of a single halide, we notice a much nicer linear trend ($R^2 \geq 0.85$), with a VBM energy ordering following the series Cl < Br < I. This small but not negligible effect is due to the energy of X n$p$ orbitals, which grows along the series just introduced. The chemical nature of the halide is more important in the case of Pb-based perovskites because the VBM orbital is more ionic than in corresponding Sn-based systems. We also notice that the slop of the linear fitting is high in both Sn and Pb-based perovskites, indicating a high correlation between $O_{VBM}$ vs $E_g$. This confirms our analysis, that all the modifications considered in experiments affect the band gap *via* the B m$s$ and X n$p$ overlap.

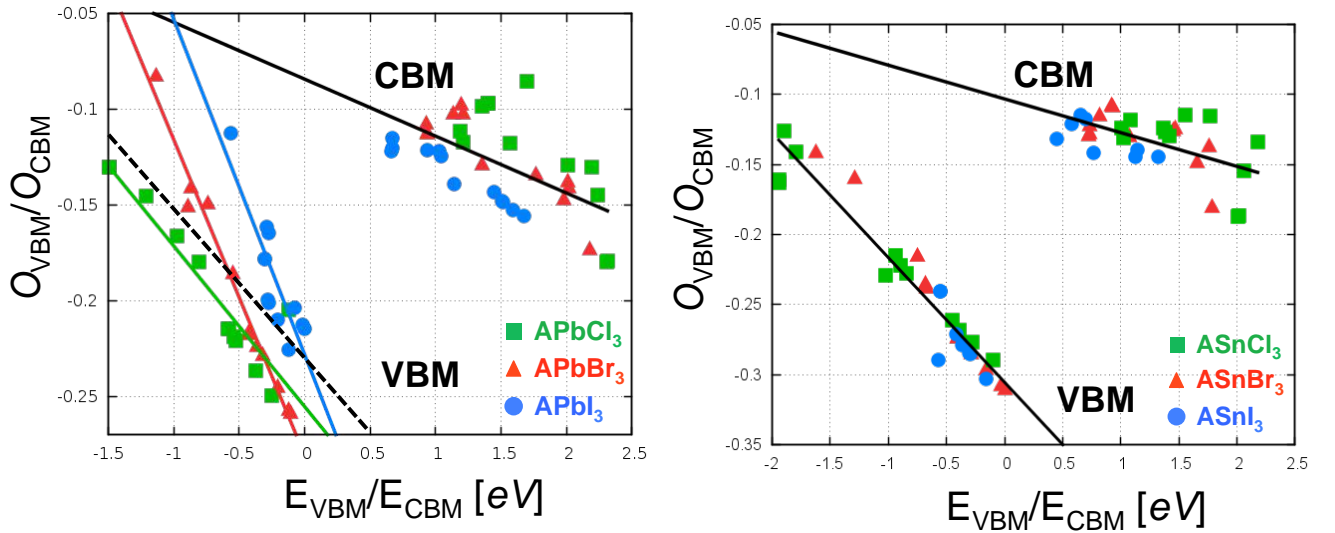

**Figure 3:** $E_{VBM}(E_{CBM})$ *vs* $O_{VBM}(O_{CBM})$ for Pb-based (right) and Sn-based (left) perovskites. $E_{BCB}$ is shifted as explained in the SI. The relation $E_{VBM}$ *vs* $O_{VBM}$ is linear ($R^2 \sim 0.93$ and $\sim 0.53$ for Sn-based and Pb-based perovskites, respectively) and has a large negative slope. In The case of Pb perovskites, if we fit $E_{VBM}$ *vs* $O_{TVB}$ independently for each halides (continuous colored lines in the left panel) we obtain a much higher $R^2$, $\geq 0.85$. The linear fitting of $E_{CBM}$ *vs* $O_{CBM}$ is poorer ($R^2 \sim 0.33$ and $\sim 0.38$ for Sn-based and Pb-based perovskites, respectively), and slope is lower.

The above argument could be used also to explain the dependency of the CBM on the chemical composition and crystal symmetry of halide perovskites. Also in this case, $O_{CBM}$ vs $E_{CBM}$ data show the expected trend of decrease of the energy of CBM with the overlap. However, data are more scattered and the slope is lower.

Indeed, in the case of CBM the change of energy is due to the electron density charge transfer from B to X (or *vice versa*) induced by the change of chemical composition and crystal structure of perovskites (see Figures 2/E and F, and SI4). Being the region around B positive, and that around X negative, moving charge from the first to the second species increases the electrostatic energy of the orbital. This explanation is proven to be correct by computing the correlation between the charge on the B ion, $q_B$, and $E_{CBM}$. Indeed, there is a very nice linear correlation between these two observables for the various perovskites of given B and X ions. The effect of changing halides from I to Br to Cl is to shift this linear correlation to higher energy values. This is because the Cl environment, which is more negative, is more repulsive to an additional electron than Br, which, in turn, is more repulsive than I.

It is remarkable that the composition and structural parameters affecting $q_B$ in the direction of increasing $E_{BCB}$ affect $O_{VBM}$ and $E_{CBM}$ in the opposite direction (Figure SI5). Thus, a single parameter, $O_{VBM}$, can be used to explain the wide experimental and computational phenomenology. The consequence of this fact is twofold. First, that chemical or structural alterations affect $E_g$ by a synergic action on $E_{VBM}$ and $E_{CBM}$: they change together in the direction of opening or closing the band gap (Figure SI1). Second, at variance with other light harvesting materials, is not easy to tune independently the energy of the valence and conduction band of perovskites to enhance charge injection in the hole and electron transport materials. This does not mean that specific changes cannot affect more one or the other band. In fact, we found modifications affecting more one or the other. Rather, we must conclude that for halide perovskites it is not possible to derive simple design principles to selectively change the energy of one or the other.

## Conclusions

We can summarize our findings as follows: chemical and structural changes increasing the negative overlap between orbitals of B and X atoms result a shrinking of the band gap. We can then use this principle for suggesting practical rules for tuning (lowering) the gap. The increase of the overlap can be achieved by choosing B and X so as to reduce the ratio between the size of the lattice and the size of their $s$ and $p$ orbitals. More practical is to use ionic radii as proxy for the lattice size, [11] and covalent radii for the size of B and X orbitals. For example, the variation of $E_g$ with halide composition follows the trend of the ratio $\alpha = r_I/r_C$, between their ionic ($r_I$) and covalent ($r_C$) radii: $\alpha_I = 1.47$; $\alpha_{Br} = 1.58$; $\alpha_{Cl} = 1.67$. As for the substitution of the monovalent cation, In the case of tetragonal and orthorhombic structures A must be large so as to reduce the tilting angles, which makes the structure as cubic-like as possible. At the same time, it should not make strong bonds with the framework atoms. Thus, cations not forming hydrogen bonding are preferable. For example, $PH_4^+$ is better than $NH_4^+$: the former is bigger than the latter, and it forms weaker hydrogen bonds.

Another remark is that tuning individual bands of perovskite to improve charge carrier transfer to hole and electron transport materials is more complex because, in general, the parameters that affect the energy of VBM also affect the energy of CBM in the opposite direction.

## Methods

Density Functional Theory (DFT) calculations are performed using the Quantum Espresso suite of codes.[13] We use the Generalized Gradient Approximation (GGA) to Density Functional Theory in the Perdew-Burke-Ernzerhof (PBE) formulation [14] and, for selected systems, the range separable hybrid exchange and correlation functional of Heyd, Scuseria, and Ernzerhof (HSE). [15] The interaction between valence electrons and core electrons and nuclei is described by ultrasoft pseudopotentials. Norm conserving pseudopotentials are used in the case of HSE calculations. In GGA calculations, Kohn-Sham orbitals are expanded on a planewave basis set with a cutoff of 40 Ry, and a cutoff of 240 Ry for the density in the case of ultrasoft pseudopotentials. HSE calculations are performed with a cutoff of 60Ry. The Brillouin zone is sampled with a 3x3x3 and 4x4x4 Monkhorst-Pack k-points grid [16] for cubic and tetragonal/orthorhombic structures, respectively. The above values are chosen by checking the convergence of total energy, band gap and atomic forces.

Sn-based computational samples are prepared starting from experimental structures for $CsSnBr_3$,[17] replacing $Cs^+$ with $Na^+$, $Li^+$, $NH_4^+$ and $PH_4^+$; and $Br^-$ is replaced with $Cl^-$ and $I^-$, as needed. Structures (atomic positions and lattice parameters) are then relaxed. Cubic samples consist of a 2x2x2 supercell containing 8 stoichi-ometric units (40 atoms). For the body-centered tetragonal structures, we considered the simple tetragonal analogue, which contains 4 stoichiometric units (20 atoms). Finally, for the orthorhombic structure we used the experimental unit cell, which already contains 4 stoichiometric units. Pb-based samples are created using the same protocol starting from experimental structures of Refs. [18] and [19].

Additional details of the computational setup, namely the effect of exact exchange, are reported in the Supp. Info.

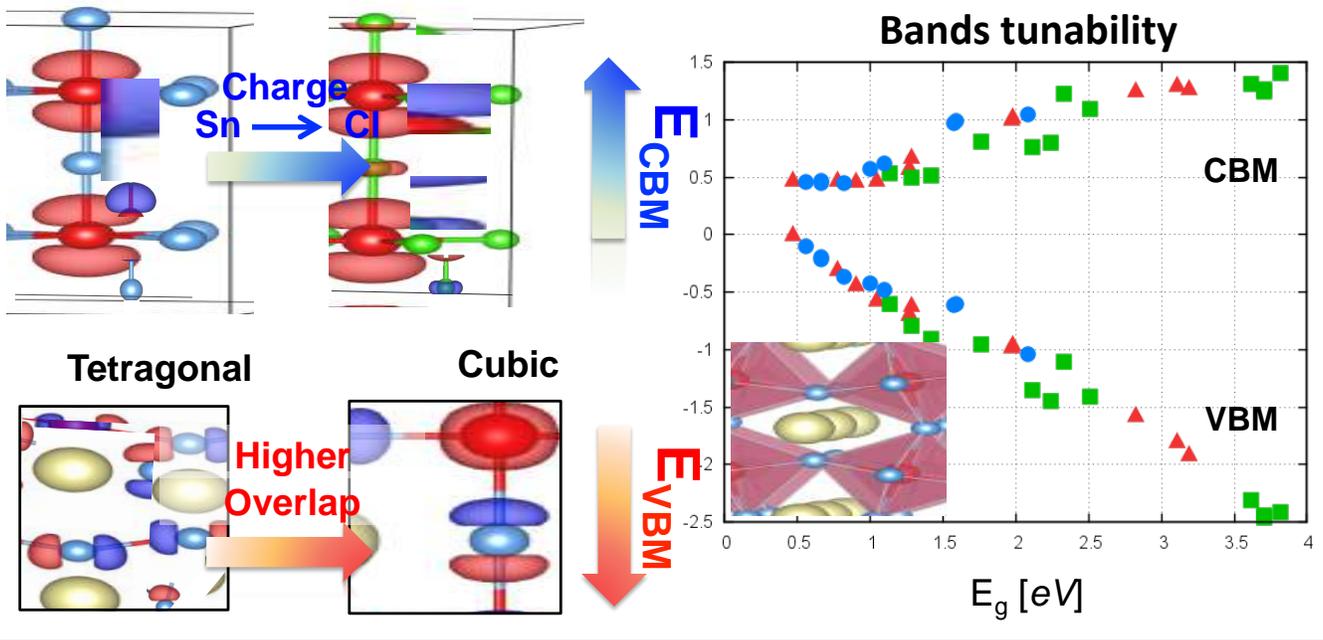

# Supplementary Information

## Additional computational details

We performed calculations with and without SOC corrections. The effect of SOC corrections to the GGA-PBE $E_g$ is, typically, of the order of 0.3-0.4 and ~0.7-1.3 eV toward smaller band gaps for Sn system and Pb-based perovskites, respectively. SOC corrections do not change the qualitative conclusions of our study: i) both VBM and CBM are important in the tuning of the band gap (Figure SI1), ii) the energy of the VBM is mostly affected by the overlap (Figure SI2), and iii) that of the CBM mostly by the charge on the B atom (Figure SI4).

In Ref. [20] it is shown that the good agreement between experimental and GGA-PBE computational results for MAPbI$_3$ is due to a cancellation of errors, with the underestimation of the band gap associated to the lack of many-body effects compensated by the overestimation associated to the lack of SOC corrections. This compensation is worse in the case of Sn-based perovskites. Here we tested the effect of a functional including the exact exchange. In fact, it might be the exact exchange in the localized VBM and CBM orbitals the main correction to the GGA results. To test this hypothesis we computed the HSE band gap, $E_g^{HSE}$, of the CsSnI3 perovskites (Cubic, tetragonal and Orthorhombic phases), added SOC corrections (described below), and compared computational and experimental results (see Figure SI3). SOC corrections are computed from the difference of GGA-PBE results with and without SOC: $\Delta E_g^{SOC} = E_g^{SOC} - E_g$. The so computed band gap of the orthorhombic CsSnI3 (1.05-1.14 eV) is in fair agreement with recent 1.2 eV experimental value. [17,21]

CBM energies of all systems are rigidly shifted so that the $E_g$ of the orthorhombic CsSnI3 is 1.14 eV, i.e. our best estimate of the band gap of this system including exact exchange and SOC effects.

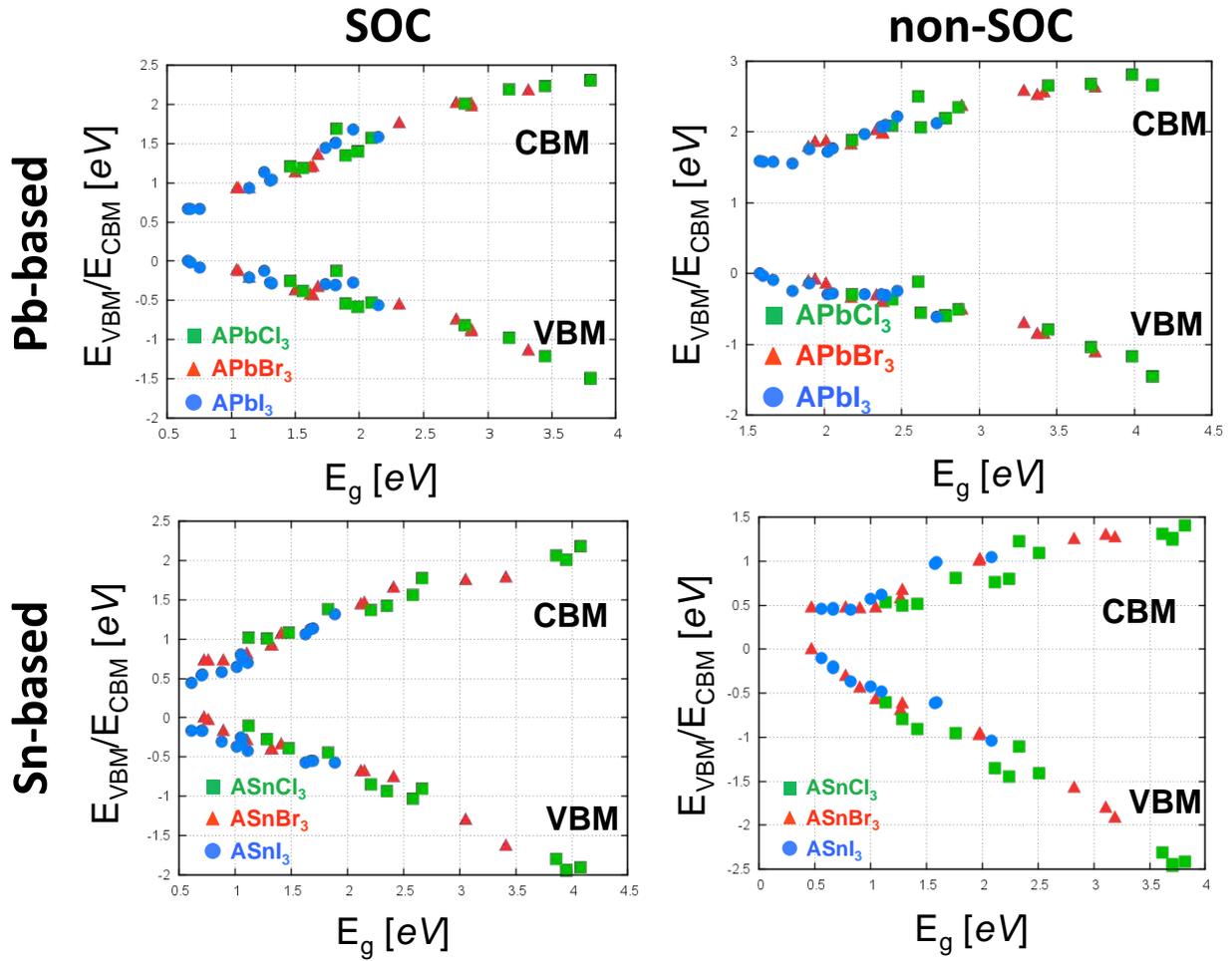

**Figure SI1:** $E_{VBM}$ and $E_{CBM}$ vs $E_g$. The top row refers to Pb-based systems, while the bottom row to Sn-based ones. On the left column we report results obtained including SOC, on the right column results obtained without SOC.

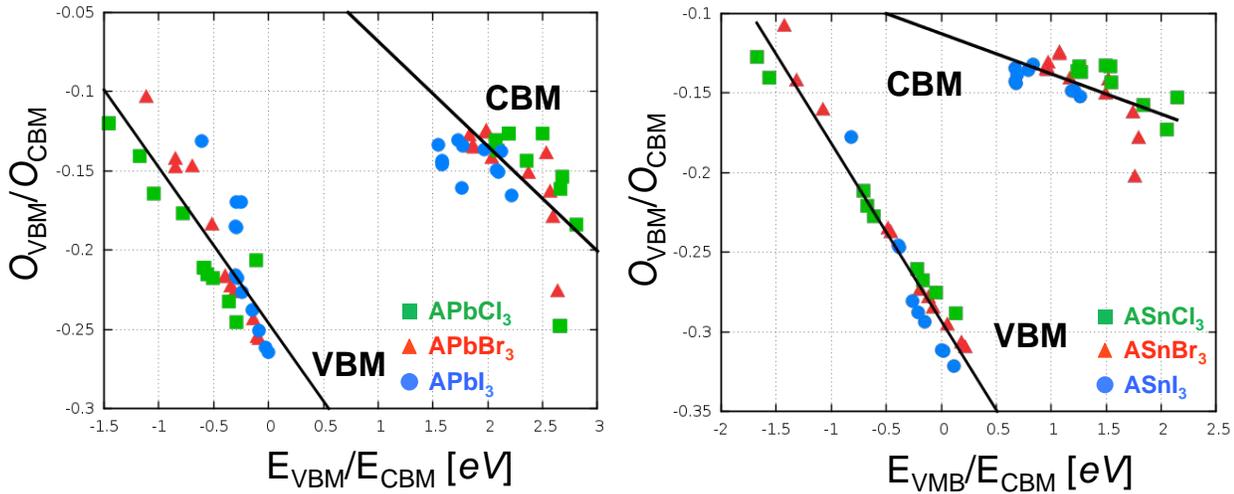

**Figure SI2 (see also Figure 3):** $E_{VBM}(E_{CBM})$ vs $O_{VBM}(O_{CBM})$ without SOC. For Sn-based systems (right), the relation $E_{VBM}$ vs $O_{VBM}$ is linear and its slope is large. $E_{CBM}$ vs $O_{CBM}$, instead, cannot be satisfactorily fit with a linear relation and presents a smaller negative slope. For the Pb-based systems (right), the linear fitting of both $E_{VBM}$ vs $O_{CBM}$ and $E_{CBM}$ vs $O_{CBM}$ is poor. At variance with the Sn-based perovskites case, here the slope of $E_{VBM}$ vs $O_{VBM}$ and $E_{CBM}$ vs $O_{CBM}$ fitting is comparable

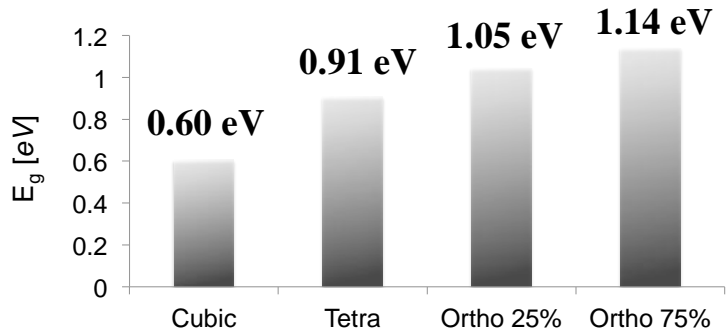

**Figure SI3:** $E_g$ **for the various crystal structures of CsSnI$_3$. Ortho 25% and Ortho 75% refer to the two orthorhombic structures identified by Kanatzidis and coworkers.** [17] **The fitting of crystallographic data required the adoption of a splitting position model for Cs and I atoms, with occupancy 25 and 75% of the two sites of each atomic species. The band gap reported is obtained from DFT calculations using the HSE exchange and correlation functional.** [15] **The SOC-corrected HSE $E_g$ (see text above) of the Ortho-rhombic 75% phase, 1.14 eV, is in nice agreement with the experimental value, 1.3 eV.** [17,21]

## Correlation between $q_B$ and $E_{CBM}$

In Figure SI4 we report $q_B$ vs $E_{CBM}$. $q_B = \sum_{i \epsilon B} c_i^2$, with $c_i$ coefficient of projection of the CBM state over B atomic orbitals, is a measure of the amount of CBM charge on B atoms. Figure SI4 shows a clear correlation between $E_{CBM}$ and $q_B$: the higher is the charge on B the lower is $E_{CBM}$. As explained in the main text, $E_{CBM}$ grows with the shrinking of $q_B$ because part of its charge is moved from the attractive $B^{2+}$ environment to the more repulsive X$^-$ one. We notice that the trend of $E_{CBM}$ with $q_B$ depends on the halide, with an almost rigid shift toward higher $E_{CBM}$ values along the series I → Br → Cl. This is not surprising because along this series the partial negative charge on the halide increases, and so does the electrostatic energy associated to the CBM charge on X. Results with and without SOC present an analogous trend, confirming that the qualitative picture one could gain from pure GGA calculations is correct.

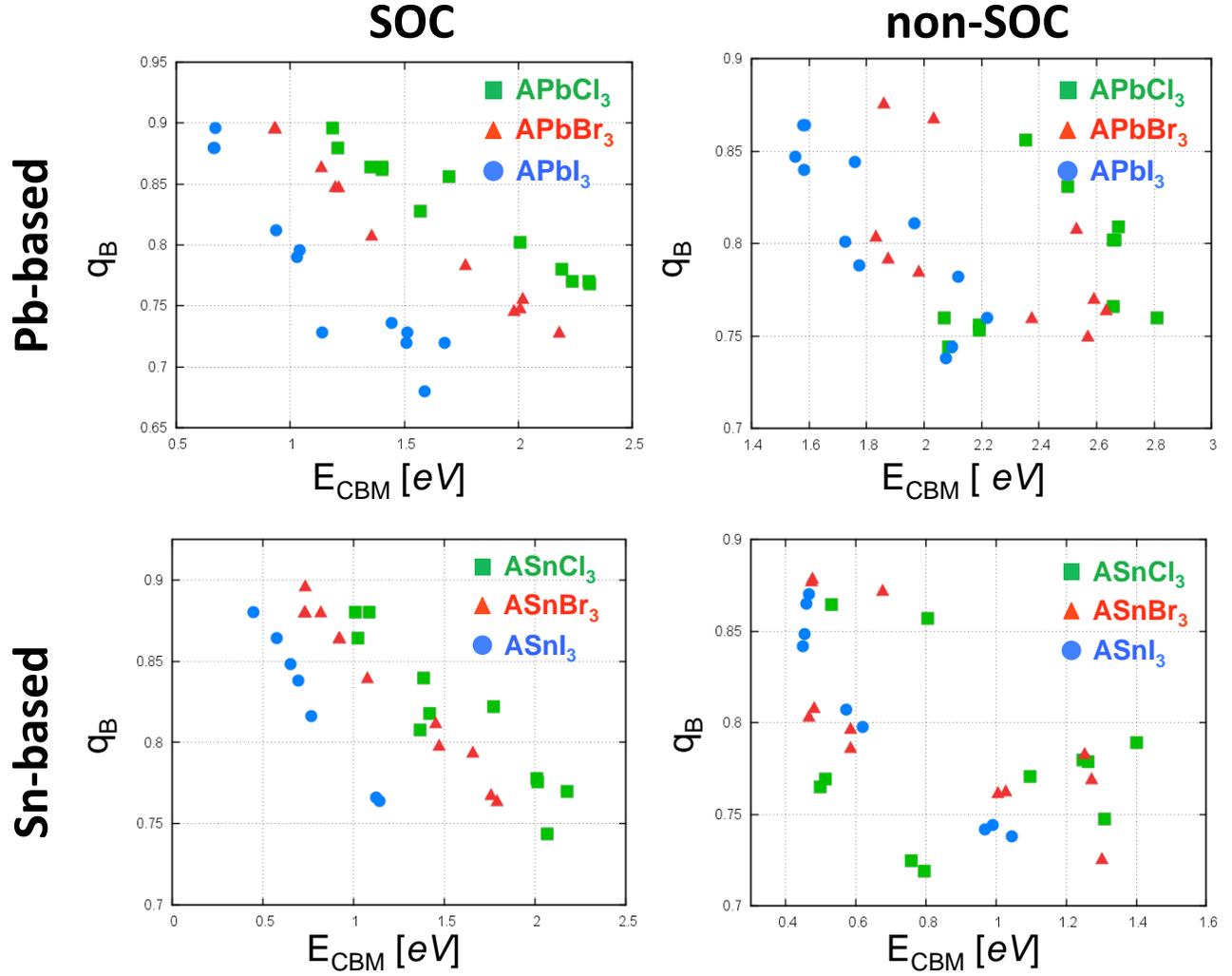

**Figure SI4:** $q_B$ **and** $E_{CBM}$. The top row refers to Sn-based systems, while the bottom row to Pb-based ones. On the left column we report results obtained including SOC, on the right column results obtained without SOC.

## Correlation between $q_B$ and $O_{TVB}$.

In the main text we explained that bands and band gap tuning strategies can be based on the overlap in the VBM crystal orbital because there is a correlation between this observable, determining the energy of the VBM, and $q_B$, determining the energy of the CBM. This is shown in Figure SI5, where we report $q_B$ vs $O_{VBM}$. We notice that $q_B$ decreases with $O_{VBM}$, which results in a cooperative action in reducing/increasing the gap: when $O_{VBM}$ changes in the sense of reducing the energy of the VBM, $q_B$ changes in the direction of increasing the energy of the CBM, and *vice versa*.

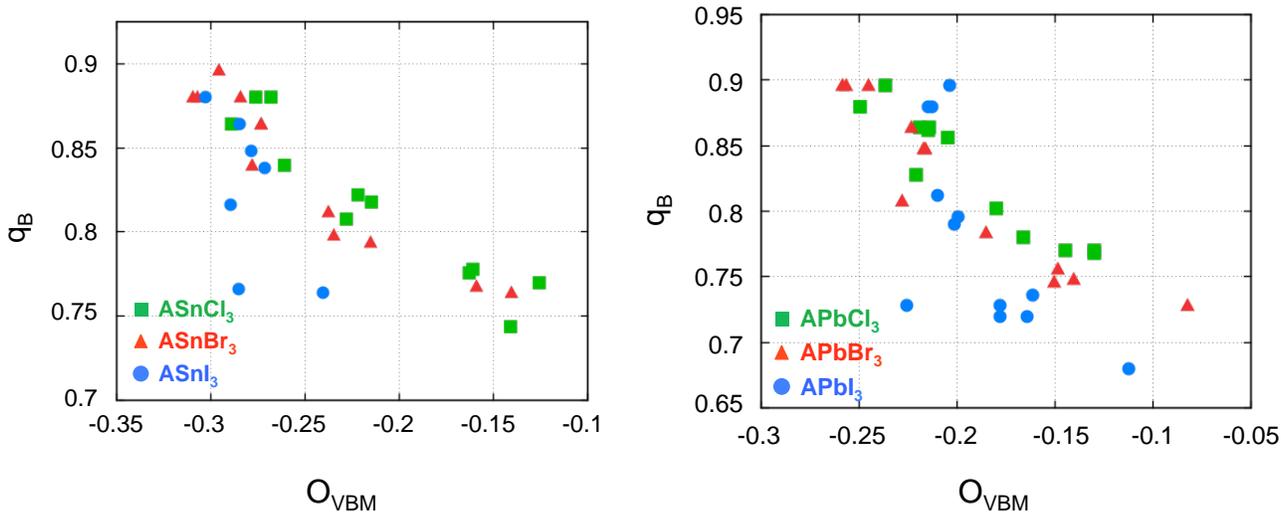

Figure SI5: $q_B$ vs $O_{VBM}$ for Sn-based (left) and Pb-based (right) perovskites.

## Calculation of the variation of $E_{VBM}$ and $E_{CBM}$ with stoichiometry and crystal symmetry, and the dependence of $E_g$ on them

Comparing the energy of Kohn-Sham (KS) orbitals in solid-state systems is difficult because there is no reference value. To overcome this problem, so as to be able to study the trend of $E_{VBM}$ and $E_{CBM}$ with chemical composition and crystal symmetry, we rigidly shifted the Kohn-Sham eigenvalues of each system of perovskites at given B so as to align low lying pure Sn/Pb states. We expect that the energy of these states, which are not directly involved in chemical bonds, is constant, and thus represent a convenient reference. Energies are further shifted so that the highest value of $E_{VBM}$ is 0 eV.

Figure SI1 shows that the energy of VBM and CBM change of ~2 and 1.5 eV in the systems investigated. Thus, we must conclude that the change of $E_g$ in the perovskites set is determined by both states. This is only apparently conflicting with the experimental results of Ref. [9]. In fact, considering the systems investigated in this latter work we also observe that the change in the band gap is mostly determined by the change in the CBM (Figure S6).

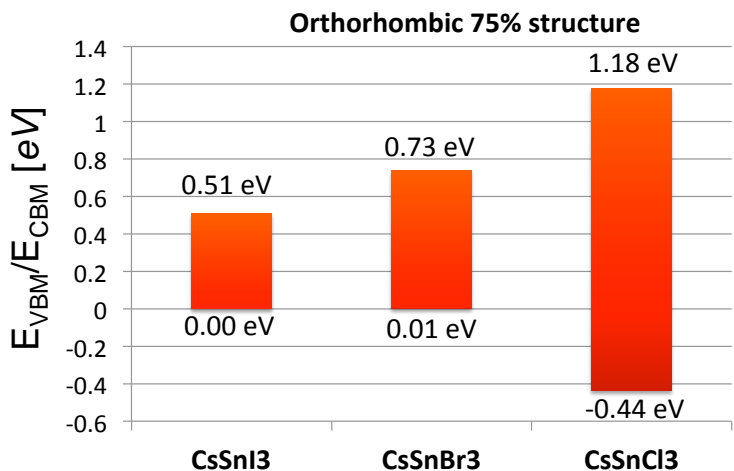

Figure SI6: $E_{VBM}$ and $E_{CBM}$ energies of the orthorhombic structure of CsSnX$_3$ perovskites.